*Review*

# Shared features of endothelial dysfunction between sepsis and its preceding risk factors (aging and chronic disease).

Jesus F Bermejo-Martin * [1,2], Marta Martín-Fernandez [1], Cristina López-Mestanza [1], Patricia Duque [3], Raquel Almansa [1,2]

1. Group for Biomedical research in Sepsis (Bio·Sepsis), Hospital Clínico Universitario de Valladolid /IECSCYL, Av. Ramón y Cajal, 3, 47003 Valladolid (Spain).

2. Centro de Investigación Biomedica En Red-Enfermedades Respiratorias (CibeRes, CB06/06/0028), Instituto de salud Carlos III (ISCIII), Av. de Monforte de Lemos, 5, 28029 Madrid (Spain).

3. Anesthesiology and Reanimation Service, Hospital General Universitario Gregorio Marañón, Madrid, Spain, Calle del Dr. Esquerdo, 46, 28007 Madrid (Spain).

Jesus F Bermejo-Martin and Marta Martín-Fernandez contributed equally to this work

* Correspondence: jfbermejo@saludcastillayleon.es; Tel.: +34 983420000 ext 87422.

**Abstract:** Acute vascular endothelial dysfunction is a central event in the pathogenesis of sepsis, increasing vascular permeability, promoting activation of the coagulation cascade, tissue edema and compromising perfusion of vital organs. Aging and chronic diseases (hypertension, dyslipidaemia, diabetes mellitus, chronic kidney disease, cardiovascular disease, cerebrovascular disease, chronic pulmonary disease, liver disease, or cancer) are recognized risk factors for sepsis. In this article we review the features of endothelial dysfunction shared by sepsis, aging and the chronic conditions preceding this disease. Clinical studies and review articles on endothelial dysfunction associated to sepsis, aging and chronic diseases published in PubMed were considered. The main features of endothelial dysfunction shared by sepsis, aging and chronic diseases were: 1) increased oxidative stress and systemic inflammation, 2) glycocalyx degradation and shedding, 3) disassembly of intercellular junctions, endothelial cell death, blood–tissue barrier disruption, 4) enhanced leukocyte adhesion and extravasation, 5) induction of a pro-coagulant and anti-fibrinolytic state. In addition, chronic diseases impair the mechanisms of endothelial reparation. In conclusion, sepsis, aging and chronic diseases induce similar features of endothelial dysfunction. The potential contribution of the pre-existent degree of endothelial dysfunction to sepsis pathogenesis deserves to be further investigated.

## 1. Introduction

Sepsis is a major health problem worldwide. Data coming exclusively from high-income countries suggests that 50.9 million cases of sepsis occur globally each year, with potentially 5.3 million deaths annually [1]. Global burden of this disease is thought to be much higher, since data on sepsis incidence in low-income and middle-income countries remain scarce. Vascular endothelial dysfunction (ED) is a central event in the pathophysiology of sepsis [2]. ED precedes organ dysfunction and plays an important role in its pathogenesis by increasing vascular permeability, promoting activation of the coagulation cascade, tissue edema and compromising regional perfusion in vital organs [3].

Aging and chronic co-morbidities are recognized risk factors of sepsis. In a recent report from the American Centers for Disease Control with 246 sepsis patients, median age was 69 yrs. Most of the patients in this study (97%) had at least one co-morbidity. 35% had diabetes mellitus, 32% had cardiovascular disease (including coronary artery disease, peripheral vascular disease, or congestive heart failure), 23% had chronic kidney disease, and 20% had chronic obstructive pulmonary disease [4]. Two large epidemiological studies on sepsis which already use the new SEPSIS-3 criteria for defining this disease provide a similar picture of the clinical characteristics of sepsis patients. The studies of Rhee *et al* with 173 690 patients [5] and Donnelly *et al* with 1080 patients [6] reveal that the mean age of sepsis patients was 66.5 yr and 69.7 yr respectively. In these large reports, the most frequent co-morbidities present in sepsis patients were those participating of the metabolic syndrome (hypertension, dyslipidaemia, diabetes mellitus), chronic kidney disease, cardiovascular disease, cerebrovascular disease, chronic pulmonary or liver disease, and cancer (Table 1). Aging and the co-morbidities preceding sepsis are characterized by inducing chronic ED. In consequence, it is probably that the acute endothelial injury induced by sepsis in aged / chronic disease patients is occurring on an endothelium which is, to a greater or lesser extent, already damaged.

| Rhee *et al* (N = 173690) | | Donnelly *et al* (N = 1080) | |
|---|---|---|---|
| Age (mean in yrs) | 66.5 | Age (mean) | 69.7 |
| Sex (male) | 57.6% | Sex (male) | 59.2% |
| Diabetes | 35.7 % | Hypertension | 74.5 % |
| Chronic pulmonary disease | 30.9 % | Dyslipidaemia | 67.3 % |
| Renal disease | 26.8 % | Diabetes | 41.8 % |
| Congestive heart failure | 25.4 % | Chronic kidney disease | 31.5 % |
| Cancer | 19.7 % | Myocardial infarction | 24.4 % |
| Dementia or cerebrovascular disease | 10.3 % | Chronic Lung disease | 17.4 % |
| Liver disease | 10 % | Stroke | 12.6 % |

**Table 1. Mean age, sex and major co-morbidities associated to sepsis**. These data correspond to the studies of Rhee et al [5] and Donnelly et al [6]. Co-morbidities are showed by their observed prevalence in each study.

This article identifies five major features of ED shared by sepsis, aging and the chronic diseases preceding this severe condition.

**2. Search strategy and selection criteria:**

References for this narrative review were identified through searches of PubMed for articles, giving priority to those published in the last 10 years, which constitutes 95% of the articles cited. Terms used were "endothelial dysfunction", "endothelium", "sepsis", "aging", "elderly", and each one of the co-morbidities associated to sepsis. Other terms searched in combination with "endothelial dysfunction" were "repair", "progenitor cells", "chemotherapy" and "radiotherapy".

## 3. The healthy endothelium

The vascular endothelium constitutes a semi-permeable barrier lining the inner surface of blood vessels (Figure 1). It controls the exchange of fluids, solutes, plasma proteins and leucocytes, by opening and closing the cell junctions composing it in a coordinated manner [7]. The normal vascular endothelium consists of a layer of endothelial cells (ECs), supported on a basement membrane, with the glycocalyx on the luminal side [8]. It prevents microorganisms to enter into tissues, exerting in addition a natural anticoagulant action that prevents from uncontrolled activation of the coagulation system.

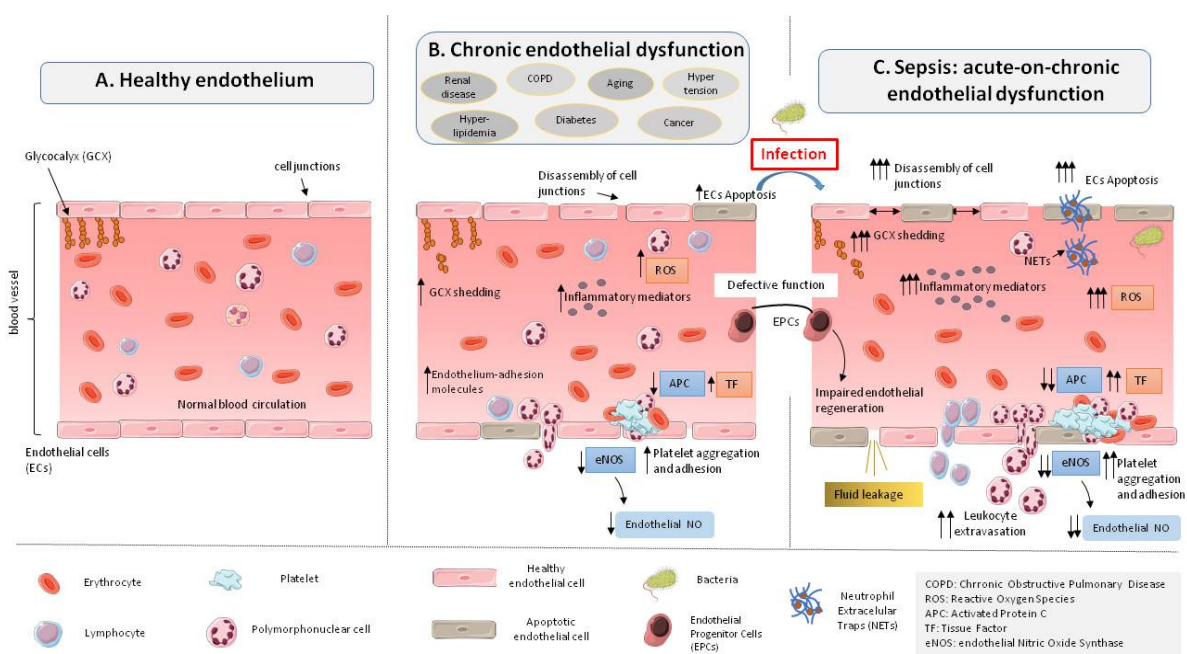

**Figure 1: The endothelium in different scenarios**: *A. Healthy endothelium*: the normal vascular endothelium consists of a layer of endothelial cells with the glycocalyx on the luminal side. It prevents microorganisms to enter into tissues, exerting in addition a natural anticoagulant action that prevents from uncontrolled activation of the coagulation system. *B. ED induced by aging and chronic disease*: senescence and the comorbidities preceding sepsis are associated to the presence of a chronic status of increased oxidative stress and inflammation, which induces glycocalyx degradation, apoptosis of endothelial cells, disassembly of endothelial cell junctions, and increased expression of molecules which promotes leukocyte adhesion to endothelial cells. In turn, these diseases induce a pro-coagulant and anti-fibrinolytic state with diminished activation of protein C and increased production of tissue factor. Decrease in the production of nitric oxyde by the endothelial nitric oxide synthase promotes platelet aggregation and adhesion. Finally, these diseases impair production and function of Endothelial Progenitor Cells, impairing endothelial regeneration. *C. ED in sepsis*: sepsis induces similar features of ED to those caused by aging and chronic diseases, inducing oxidative stress and inflammation, release of NETs and proteases by neutrophils, leading to fluid and cell leakage, hypotension, microvascular thrombosis, inadequate organ perfusion, organ failure and shock in the most severe cases. In addition, bacterial toxins can breach the endothelial barrier, by directly killing ECs, weakening the cytoskeleton within ECs, and breaking the junctions between ECs. Acute challenges such as the aggression induced by surgery, trauma or hypervolemia could contribute to facilitate or enhance ED in patients facing an infection. Images for representing cells were taken from "Smart Servier Medical Art" (https://smart.servier.com/).

*3.1 Glycocalyx (GCX):* it is an organized layer of sulfated proteoglycans, hyaluronan, glycoproteins, and plasma proteins that adhere to a surface matrix which coats the luminal surface of the endothelium. It serves as a protective barrier between the flowing blood and the vessel wall, contributing to maintain the endothelial barrier to fluid and protein, to regulate leukocyte-endothelial adhesion and to inhibit intravascular thrombosis [9].

*3.2 Endothelial cells (ECs):* ECs line our vasculature, as a one continuous layer resting on a basement membrane formed by collagen, laminins, nidogens/entactins, and the proteoglycan perlecan. Endothelial cells lining the vessel wall are connected by adherens junctions (mainly composed of vascular endothelial (VE)-cadherin, tight junctions (predominantly consisting of occludins and claudins) and gap junctions [2] [10], which prevent leukocyte emigration and vascular leak [2]. Embedded in the basement membrane and outside it is a non-continuous layer of cells known as pericytes, which are thought to play a role in angiogenesis [8].

## 4. ED in sepsis

Sepsis causes acute ED, inducing a pro-adhesive, pro-coagulant and anti-fibrinolytic state in ECs, altering hemostasis, leukocyte trafficking, inflammation, barrier function, and microcirculation [11] (Figure 1).

*4.1 Increased oxidative stress and systemic inflammation*: there are a number of mediators participating in the "molecular storm" occurring in sepsis that initiate and amplify injury to the endothelium. Between these molecules are bacterial endotoxins / pathogen-associated molecular patterns (PAMPs), cytokines, bradykinin, histamine, platelet-activating factor (PAF), vascular endothelial growth factor (VEGF), fibrin degradation products and reactive oxygen species (ROS) such as hydrogen peroxide, hydroxyl anions, and superoxide [3] [12] [13] [14]. In turn, the endothelium it is not just a passive element suffering the aggression during sepsis, but it also produces chemokines to attract immune cells, boosting the inflammatory response [14].

*4.2 GCX degradation and shedding*: the "cocktail" of pro-inflammatory and pro-oxydative molecules induced by sepsis promotes shedding of the GCX [2] [3] [15]. Release of damage- associated molecular patterns (DAMPs) such as degradation products of the endothelial GCX (i.e heparan sulfates) or components of Neutrophil extracellular traps (NETs) amplifies this deleterious response [9] [15].

*4.3 Disassembly of intercellular junctions, endothelial cell death, blood–tissue barrier disruption*: the marked pro-inflammatory and oxidative response occurring in sepsis induces also the formation of gaps between ECs by disassembly of intercellular junctions [2] [3] [15]. NETs released from dying neutrophils induce death of ECs, an effect mediated by NETs-related proteases and cationic proteins such as defensins and histones [11] [16]. Bacterial toxins can breach the endothelial barrier, by directly killing ECs, weakening the cytoskeleton within ECs, and breaking the junctions between ECs [13].

*4.4 Enhanced leukocyte adhesion and extravasation*: Glycocalyx shedding exposes the endothelium to leukocyte adhesion [15]. Pro-inflammatory cytokines induce expression of molecules such as P-selectin, E-selectin, ICAM1 or VCAM1 that allow adhesion of activated immune cells to the vascular wall and promote transendothelial migration into surrounding tissues [3]. Activated neutrophils from sepsis patients adhered to endothelium mediate profound loss of endothelial barrier integrity [17]. The proteases released by activated neutrophils could contribute to degrade junctional proteins [8]. Extravased neutrophils induce tissue damage producing potentially destructive enzymes and oxygen-free radicals.

*4.5 Induction of a pro-coagulant and anti-fibrinolytic state*: in sepsis there is also a global increase in the production of nitric oxide (NO, a potent vasodilator) mediated by the inducible nitric oxide synthase (iNOS) [3] [18]. In contrast, there is an important decrease in the production of NO by endothelial nitric oxide synthase (eNOS), which impairs direct vasodilatation, and promotes platelet and leukocyte adhesion [12]. Down-regulation of endothelial expression of thrombomodulin and endothelial protein C receptors translates into diminished activation of the activated protein C [11]. ECs release the procoagulant glycoprotein TF (tissue factor), whereas their synthesis of TF pathway inhibitor is inhibited [11]. The activation of platelets & the coagulation cascade causes microvascular thrombosis [9]. In addition, NETs provide a scaffold for thrombus formation, promoting hypercoagulability in patients with sepsis [11]. The association of TF with NETs could target thrombin generation and fibrin clot formation at sites of infection/neutrophil activation, with active thrombin leading to increased platelet activation [19]. Acute vascular dysfunction and leakage contribute to hypotension, inadequate organ perfusion, local hypoxia, ischemia and ultimately, to organ failure, acute respiratory distress syndrome, shock and death in the most severe patients [12]

**5. ED associated to aging and chronic disease.**

The same features of ED induced by sepsis are also induced by aging and chronic disease (Figure 1):

*5.1 Increased oxidative stress and systemic inflammation*: aging is characterized by the presence of increased endothelial oxidative stress, as a result of augmented production from the intracellular enzymes NADPH oxidase and uncoupled eNOS, as well as from mitochondrial respiration in the absence of appropriate increases in antioxidant defenses [20]. Nitroso-oxydative stress contributes to ED associated with diabetes [21]. Vascular oxidative stress and inflammation are major determinants of ED in atherosclerosis and cardiovascular diseases [22]. Inflammation and free radical formation contributes also to the pathogenesis of hypertension and cancer [23]. In patients with Chronic Obstructive Pulmonary Disease (COPD), chronic inflammation does not impact only on lung parenchyma, but potentially involves the endothelium of blood vessels, which makes it a systemic disease [24]. Inflammation and oxidative stress play major roles in the pathogenesis of ED in liver cirrhosis [25].

*5.2 GCX degradation and shedding*: untreated hypertension, diabetes mellitus and hypercholesterolemia are associated to a reduced endothelial GCX thickness [26] [27] [28]. Hyperglycaemia and oxidised low-density lipoprotein (LDL) causes GCX dysfunction [29]. GCX degradation is an initial event in atherosclerosis, promoting lipid deposition in the vessel wall [30]. Patients with chronic renal disease under dialysis have an impaired GCX barrier and shed its constituents into blood [31]. Cigarettes smoking (the main cause of COPD) compromises endothelial GCX integrity [32]. Patients with end-stage liver disease show marked increased concentration of syndecan-1 in plasma, a marker of GCX shedding [33]. Elevated haematocrit, a risk factor for stroke and myocardial infarction, could induce a reduction in GCX thickness [34]. Patients with heart failure with reduced ejection fraction have increased levels of the GCX shedding marker median hyaluronic acid [35]. Lacunar stroke patients with white matter lesions show compromised GCX barrier properties [36].

*5.3 Disassembly of intercellular junctions, endothelial cell death, blood–tissue barrier disruption*: sedentary aging enhances endothelial cell senescence. Senescent ECs show a pro-oxidant phenotype with increased ROS production, which promote endothelial injury [20]. Tight junction structure and barrier integrity is significantly impaired in senescent ECs [37]. LDL from patients with hypercholesterolaemia are inflammatory to microvascular endothelial cells, impairing in addition endothelial tight junction expression [38]. Activation of endothelial inflammasomes due to increased free fatty acids produces high mobility group box protein-1, HMGB1, which disrupts inter-endothelial junctions and increases paracellular permeability of endothelium [39]. High glucose concentrations induce disruption of endothelial adherens junctions mediated by protein kinase C-β-dependent vascular endothelial cadherin tyrosine phosphorylation [40]. Patients with chronic kidney diseases have increased levels of anti-endothelial cell antibodies, and decreased expression of both adherens and tight junction proteins VE-cadherin, claudin-1, and zonula occludens-1 [41]. Cigarette smoking disrupts intercellular adhesion molecules between ECs and induces their apoptosis [42]. In COPD patients, circulating anti-endothelial cell antibodies along with chronic oxidative and inflammatory stress induces apoptosis of ECs [43]. Cancer induces also disruption of endothelial junctions, in particular adherens junctions [44]. Anticancer chemotherapy may induce systemic damage of vascular endothelium related to massive cell loss and alterations of endothelial function [45]. Radiotherapy causes premature senescence, apoptosis of endothelial cells and increased vascular permeability [46] . Endothelial barrier is also altered in cardiovascular disease. Chemical modification of tubulin caused by cardiometabolic risk factors and oxidative stress leads to reorganization of endothelial microtubules, destabilizing vascular integrity and increasing permeability, which finally results in increasing cardiovascular and cerebrovascular risk [47] . Intravascular albumin is important to maintain vascular integrity, since it contributes to preserve normal capillary permeability [48] and the GCX structure [49]. Patients with malnutrition, liver disease or nephrotic syndrom could present hypo-albuminemia.

*5.4 Enhanced leukocyte adhesion and extravasation*: aging is associated to stiffening of extracellular matrix within the intima, which promotes EC permeability and leukocyte extravasation [50]. Hypertension induces vascular wall injury and remodeling, a process which involves recruitment of leukocytes to the endothelium [23]. Glycocalix degradation in atherosclerosis facilitates the interaction between ECs and leukocytes [30]. Hyperglycemia induces activation of NF-κB in ECs, leading to an increased production of adhesion molecules, leukocyte-attracting chemokines and cytokines activating inflammatory cells in the vascular wall [51] . In chronic kidney disease, leukocytes acquire an adhesive phenotype, by mechanisms mediated by hypoxia and by cytokines released from ischemic renal endothelium [52]. In patients with COPD, fibrinogen is increased and stimulates the adhesion of platelets and white blood cells to the vessel wall [24]. Tobacco nicotine causes a loss of functional integrity of endothelium by causing vasospasm, stimulating the adhesion of leukocytes [53]. Up-regulation of selectins seems to be a central event in metastatic progression in cancer, proteins which mediate tethering and rolling of leukocytes to the vascular endothelium. Regarding cancer treatment, radiotherapy leads to increased endothelial cell activation and expression of VCAM-1, ICAM-1, PECAM-1, E-selectin and P-selectin, which promotes adhesion of leukocytes [46]. Chronic liver disease is characterized by upregulation of endothelium-adhesion molecules such as CD11b in circulating leukocytes in blood [54]. Leukocyte activation, adhesion and accumulation in the endothelium are events playing an important role in the pathogenesis of different cardiac diseases (myocarditis, cardiomyopathy, cardiac hypertrophy and failure, and ischemic heart disease) [55] and in ischemic cerebrovascular disease [56].

*5.5 Induction of a pro-coagulant and anti-fibrinolytic state*: senescent ECs show reduced eNOS activity, which impairs their ability to inhibit platelet aggregation [57]. Hypertension is associated to ED leading to attenuated NO formation because of direct oxidative modification of eNOS [58]. GCX degradation in atherosclerosis causes ECs to reduce their expression of eNOS [30]. Oxidative stress associated with hyperglicemia induces eNOS uncoupling and reduce NO production [21]. Hyperglycemia, excess free fatty acid release and insulin resistance in diabetes mellitus induces platelet hyperactivity [51]. Hyperglycemia generates also a prothrombotic state by the increased production of lesion-based coagulants, such as tissue factor, and the inhibitors of fibrinolysis, such as PAI-1 [51]. Patients with chronic renal disease can show either coagulation defects and endothelial cell damage leading to a thrombophilic state, which is characterized by the presence of high plasma concentration of fibrinogen, D-dimer, thrombin–antithrombin complex, coagulation factor VII, vWF, thrombomodulin and PAI-1[59]. Patients with COPD or liver cirrhosis have impaired eNOS activity also [24] [25]. Smoking along with the mantained pro-inflammatory state in COPD induce a thrombotic effect, by increasing platelet activation and triggering coagulation cascade [24]. Cancer may result in activation of coagulation and endothelial cell perturbation, leading to coagulopathies, a prothrombotic state and microvascular dysfunction, by mechanisms involving tissue factor-mediated thrombin generation, down regulation of endothelial cell-associated physiological anticoagulant pathways, deranged fibrinolysis and dysfunctional ECs [60]. This pro-coagulating phenotype in cancer could be favored also by chemo and radiotherapy [46] [60]. Patients with chronic liver disease show hyperhomocysteinaemia, a disorder of methionine metabolism, which has been suggested to cause endothelial injury and atherothrombotic vascular disease by several mechanisms involving oxidative stress, altered production of NO and impaired platelet-modulating activity [61]. Coronary heart disease associated with hypertension is characterized by reduced endothelial NO synthesis [62]. Endothelial dysfunction and coagulation are also involved in the pathogenesis of ischaemic stroke [63].

**6. Impact of aging and chronic disease on the mechanisms of endothelial repair**

Endothelial cell injury is mitigated by the reparative activity of bone marrow-derived endothelial progenitor cells (EPCs). As recently described, EPCs could have a role in determining the prognosis of patients with sepsis [64]. Chronic diseases preceding sepsis decrease EPC availability and/or mobilization. Cell senescence impairs the regenerative capacity of ECs. Age impairs migration of endothelial progenitor cells (EPCs) reducing their ability to contribute to vascular repair [65]. The absolute number, or functional capacity of EPCs, has been shown to be reduced in several disease states including diabetes, hypertension, hypercholesterolemia, and chronic kidney disease [66]. Circulating EPCs from COPD are dysfunctional, displaying impaired angiogenic ability and increased apoptosis [67]. Radiotherapy lower the number of circulating EPCs in cancer survivors [68]. The consequences of chemotherapy on EPCs at the long term are unknown.

**7. Discussion**

The evidence available from the literature shows that sepsis and their preceding risk factors (aging and chronic diseases) share common features of ED. In addition, aging and chronic diseases impair the regenerative ability of the endothelium. All of this draws a scenario of "endothelial frailty" preceding sepsis. Most of the features of ED are redundant in aging and the different chronic co-morbidities preceding sepsis, which could exert synergistic effects impairing endothelial function before and during sepsis. In addition, acute challenges such as the aggression induced by surgery, trauma or hypervolemia [69] [70] could contribute to facilitate or to enhance ED in patients facing an infection. In consequence, a phenomenon of acute on chronic ED could be occurring in sepsis, similar to the "acute on chronic kidney failure" of those patients with chronic renal disease (Figure 1).

A limitation of our review is that it was focused in those co-morbidities linked to sepsis in high income countries. In low-middle income countries, the epidemiological profile of sepsis is different. Sepsis patients are younger in these countries, but they suffer frequently from a number of conditions also linked to ED, such as malnutritition, malaria and HIV infection. Another limitation is that our review addresses only vascular ED. There is a total lack of information in the literature on the impact of sepsis, aging and chronic disease on the lymphatic vessels endothelium. The lymphatic system could have important implications in sepsis physiopathology, since it is the main avenue for circulation of dendritic cells and lymphocytes. Finally, studies evaluating the interactions between chronic endothelial, immunological, coagulation and metabolic dysfunction will help to better understand the pathogenesis of sepsis (Figure 2).

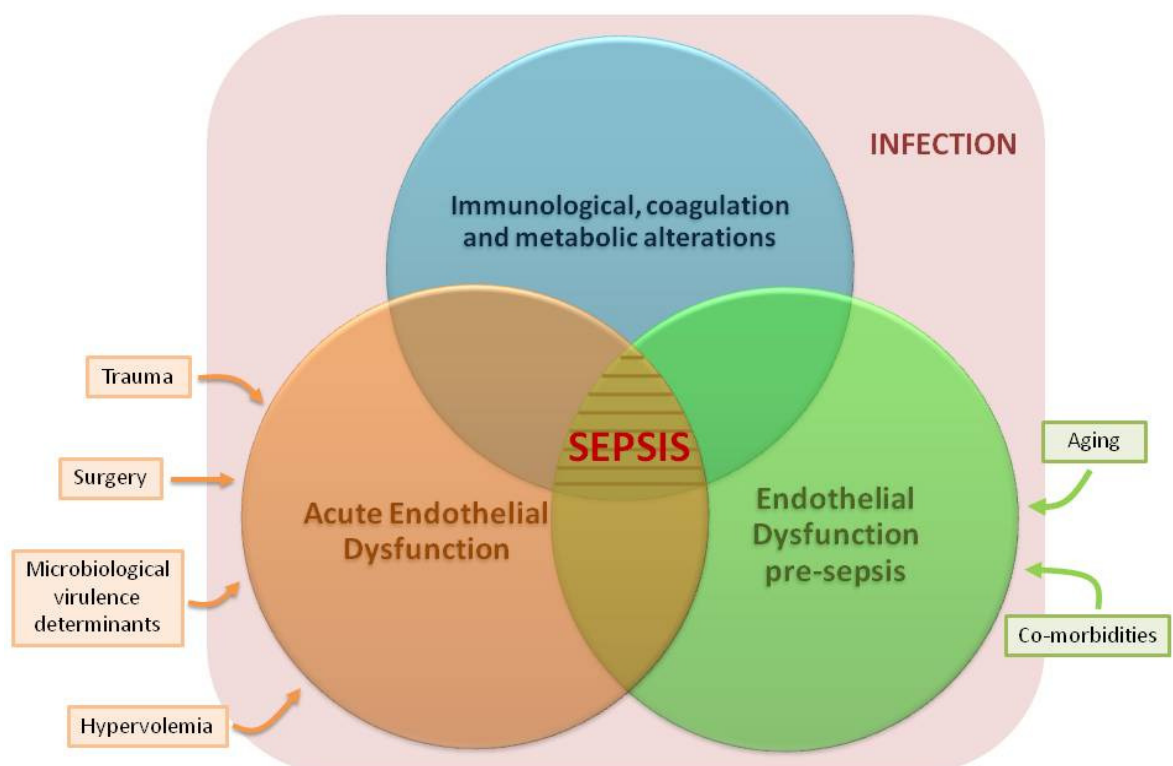

**Figure 2. Potentially synergistic factors contributing to sepsis.** *Blue circle*: Immunological, coagulation or metabolic alterations previous to sepsis or induced by sepsis. *Green circle*: Chronic ED pre-sepsis caused by aging and chronic diseases. *Orange circle*: Acute ED induced by sepsis secondary to microbial aggression, to the dysfunctional host response to infection, hypervolemia, trauma, surgery, and/or to other acute insults facilitating sepsis.

## 8. Conclusions

Sepsis, aging and chronic disease induce similar features of ED. The degree of ED already present prior to sepsis could have an important role in the pathogenesis of this disease.

**Author Contributions:** The bibliographic review was performed by JFBM, who wrote the manuscript. CLM, PD, RA contributed to design the article structure, and to select the included bibliography. In addition, they critically reviewed the manuscript. MMF and RA created the figures and the tables and critically reviewed the manuscript.

**Acknowledgments:** BioSepsis is granted by "Consejería de Sanidad de Castilla y León-IECSCYL" and the "Instituto de Salud Carlos III", Spain, EMER07/050 / PI16/01156. Consejería de Educación de Castilla y Leon/Fondo social Europeo" supports the contract of Marta Martin-Fernandez. We thank the Anesthesiology and Critical Care Services of our Hospitals for their continued support to our research program in sepsis.

**Conflicts of Interest:** The authors declare no conflict of interest.

**References**

1. Hotchkiss, R. S.; Moldawer, L. L.; Opal, S. M.; Reinhart, K.; Turnbull, I. R.; Vincent, J.-L. Sepsis and septic shock. *Nat. Rev. Dis. Primer* **2016**, *2*, 16045, doi:10.1038/nrdp.2016.45.

2. Lee, W. L.; Slutsky, A. S. Sepsis and endothelial permeability. *N. Engl. J. Med.* **2010**, *363*, 689–691, doi:10.1056/NEJMcibr1007320.

3. Ince, C.; Mayeux, P. R.; Nguyen, T.; Gomez, H.; Kellum, J. A.; Ospina-Tascón, G. A.; Hernandez, G.; Murray, P.; De Backer, D.; ADQI XIV Workgroup THE ENDOTHELIUM IN SEPSIS. *Shock Augusta Ga* **2016**, *45*, 259–270, doi:10.1097/SHK.0000000000000473.

4. Novosad, S. A.; Sapiano, M. R. P.; Grigg, C.; Lake, J.; Robyn, M.; Dumyati, G.; Felsen, C.; Blog, D.; Dufort, E.; Zansky, S.; Wiedeman, K.; Avery, L.; Dantes, R. B.; Jernigan, J. A.; Magill, S. S.; Fiore, A.; Epstein, L. Vital Signs: Epidemiology of Sepsis: Prevalence of Health Care Factors and Opportunities for Prevention. *MMWR Morb. Mortal. Wkly. Rep.* **2016**, *65*, 864–869, doi:10.15585/mmwr.mm6533e1.

5. Rhee, C.; Dantes, R.; Epstein, L.; Murphy, D. J.; Seymour, C. W.; Iwashyna, T. J.; Kadri, S. S.; Angus, D. C.; Danner, R. L.; Fiore, A. E.; Jernigan, J. A.; Martin, G. S.; Septimus, E.; Warren, D. K.; Karcz, A.; Chan, C.; Menchaca, J. T.; Wang, R.; Gruber, S.; Klompas, M.; CDC Prevention Epicenter Program Incidence and Trends of Sepsis in US Hospitals Using Clinical vs Claims Data, 2009-2014. *JAMA* **2017**, *318*, 1241–1249, doi:10.1001/jama.2017.13836.

6. Donnelly, J. P.; Safford, M. M.; Shapiro, N. I.; Baddley, J. W.; Wang, H. E. Application of the Third International Consensus Definitions for Sepsis (Sepsis-3) Classification: a retrospective population-based cohort study. *Lancet Infect. Dis.* **2017**, *17*, 661–670, doi:10.1016/S1473-3099(17)30117-2.

7. Radeva, M. Y.; Waschke, J. Mind the gap: mechanisms regulating the endothelial barrier. *Acta Physiol. Oxf. Engl.* **2018**, *222*, doi:10.1111/apha.12860.

8. Gane, J.; Stockley, R. Mechanisms of neutrophil transmigration across the vascular endothelium in COPD. *Thorax* **2012**, *67*, 553–561, doi:10.1136/thoraxjnl-2011-200088.

9. Colbert, J. F.; Schmidt, E. P. Endothelial and Microcirculatory Function and Dysfunction in Sepsis. *Clin. Chest Med.* **2016**, *37*, 263–275, doi:10.1016/j.ccm.2016.01.009.

10. Wallez, Y.; Huber, P. Endothelial adherens and tight junctions in vascular homeostasis, inflammation and angiogenesis. *Biochim. Biophys. Acta* **2008**, *1778*, 794–809, doi:10.1016/j.bbamem.2007.09.003.

11. Hattori, Y.; Hattori, K.; Suzuki, T.; Matsuda, N. Recent advances in the pathophysiology and molecular basis of sepsis-associated organ dysfunction: Novel therapeutic implications and challenges. *Pharmacol. Ther.* **2017**, *177*, 56–66, doi:10.1016/j.pharmthera.2017.02.040.

12. Pool, R.; Gomez, H.; Kellum, J. A. Mechanisms of Organ Dysfunction in Sepsis. *Crit. Care Clin.* **2018**, *34*, 63–80, doi:10.1016/j.ccc.2017.08.003.

13. Lubkin, A.; Torres, V. J. Bacteria and endothelial cells: a toxic relationship. *Curr. Opin. Microbiol.* **2017**, *35*, 58–63, doi:10.1016/j.mib.2016.11.008.

14. Chousterman, B. G.; Swirski, F. K.; Weber, G. F. Cytokine storm and sepsis disease pathogenesis. *Semin. Immunopathol.* **2017**, *39*, 517–528, doi:10.1007/s00281-017-0639-8.

15. Martin, L.; Koczera, P.; Zechendorf, E.; Schuerholz, T. The Endothelial Glycocalyx: New Diagnostic and Therapeutic Approaches in Sepsis. *BioMed Res. Int.* **2016**, *2016*, 3758278, doi:10.1155/2016/3758278.

16. Saffarzadeh, M.; Juenemann, C.; Queisser, M. A.; Lochnit, G.; Barreto, G.; Galuska, S. P.; Lohmeyer, J.; Preissner, K. T. Neutrophil extracellular traps directly induce epithelial and endothelial cell death: a predominant role of histones. *PloS One* **2012**, *7*, e32366, doi:10.1371/journal.pone.0032366.

17. Fox, E. D.; Heffernan, D. S.; Cioffi, W. G.; Reichner, J. S. Neutrophils from critically ill septic patients mediate profound loss of endothelial barrier integrity. *Crit. Care Lond. Engl.* **2013**, *17*, R226, doi:10.1186/cc13049.

18. Fortin, C. F.; McDonald, P. P.; Fülöp, T.; Lesur, O. Sepsis, leukocytes, and nitric oxide (NO): an intricate affair. *Shock Augusta Ga* **2010**, *33*, 344–352, doi:10.1097/SHK.0b013e3181c0f068.

19. Gardiner, E. E.; Andrews, R. K. Neutrophil extracellular traps (NETs) and infection-related vascular dysfunction. *Blood Rev.* **2012**, *26*, 255–259, doi:10.1016/j.blre.2012.09.001.

20. Donato, A. J.; Morgan, R. G.; Walker, A. E.; Lesniewski, L. A. Cellular and molecular biology of aging endothelial cells. *J. Mol. Cell. Cardiol.* **2015**, *89*, 122–135, doi:10.1016/j.yjmcc.2015.01.021.

21. Goligorsky, M. S. Vascular endothelium in diabetes. *Am. J. Physiol. Renal Physiol.* **2017**, *312*, F266–F275, doi:10.1152/ajprenal.00473.2016.

22. Daiber, A.; Steven, S.; Weber, A.; Shuvaev, V. V.; Muzykantov, V. R.; Laher, I.; Li, H.; Lamas, S.; Münzel, T. Targeting vascular (endothelial) dysfunction. *Br. J. Pharmacol.* **2017**, *174*, 1591–1619, doi:10.1111/bph.13517.

23. Konukoglu, D.; Uzun, H. Endothelial Dysfunction and Hypertension. *Adv. Exp. Med. Biol.* **2017**, *956*, 511–540, doi:10.1007/5584_2016_90.

24. Malerba, M.; Nardin, M.; Radaeli, A.; Montuschi, P.; Carpagnano, G. E.; Clini, E. The potential role of endothelial dysfunction and platelet activation in the development of thrombotic risk in COPD patients. *Expert Rev. Hematol.* **2017**, *10*, 821–832, doi:10.1080/17474086.2017.1353416.

25. Vairappan, B. Endothelial dysfunction in cirrhosis: Role of inflammation and oxidative stress. *World J. Hepatol.* **2015**, *7*, 443–459, doi:10.4254/wjh.v7.i3.443.

26. Ikonomidis, I.; Voumvourakis, A.; Makavos, G.; Triantafyllidi, H.; Pavlidis, G.; Katogiannis, K.; Benas, D.; Vlastos, D.; Trivilou, P.; Varoudi, M.; Parissis, J.; Iliodromitis, E.; Lekakis, J. Association of impaired endothelial glycocalyx with arterial stiffness, coronary microcirculatory dysfunction, and abnormal myocardial deformation in untreated hypertensives. *J. Clin. Hypertens. Greenwich Conn* **2018**, doi:10.1111/jch.13236.

27. Groen, B. B. L.; Hamer, H. M.; Snijders, T.; van Kranenburg, J.; Frijns, D.; Vink, H.; van Loon, L. J. C. Skeletal muscle capillary density and microvascular function are compromised with aging and type 2 diabetes. *J. Appl. Physiol. Bethesda Md 1985* **2014**, *116*, 998–1005, doi:10.1152/japplphysiol.00919.2013.

28. Meuwese, M. C.; Mooij, H. L.; Nieuwdorp, M.; van Lith, B.; Marck, R.; Vink, H.; Kastelein, J. J. P.; Stroes, E. S. G. Partial recovery of the endothelial glycocalyx upon rosuvastatin therapy in patients with heterozygous familial hypercholesterolemia. *J. Lipid Res.* **2009**, *50*, 148–153, doi:10.1194/jlr.P800025-JLR200.

29. Drake-Holland, A. J.; Noble, M. I. The important new drug target in cardiovascular medicine--the vascular glycocalyx. *Cardiovasc. Hematol. Disord. Drug Targets.* **2009**, *9*, 118–123.

30. Mitra, R.; O'Neil, G. L.; Harding, I. C.; Cheng, M. J.; Mensah, S. A.; Ebong, E. E. Glycocalyx in Atherosclerosis-Relevant Endothelium Function and as a Therapeutic Target. *Curr. Atheroscler. Rep.* **2017**, *19*, 63, doi:10.1007/s11883-017-0691-9.

31. Vlahu, C. A.; Lemkes, B. A.; Struijk, D. G.; Koopman, M. G.; Krediet, R. T.; Vink, H. Damage of the endothelial glycocalyx in dialysis patients. *J. Am. Soc. Nephrol. JASN* **2012**, *23*, 1900–1908, doi:10.1681/ASN.2011121181.

32. Ikonomidis, I.; Marinou, M.; Vlastos, D.; Kourea, K.; Andreadou, I.; Liarakos, N.; Triantafyllidi, H.; Pavlidis, G.; Tsougos, E.; Parissis, J.; Lekakis, J. Effects of varenicline and nicotine replacement therapy on arterial elasticity, endothelial glycocalyx and oxidative stress during a 3-month smoking cessation program. *Atherosclerosis* **2017**, *262*, 123–130, doi:10.1016/j.atherosclerosis.2017.05.012.

33. Schiefer, J.; Lebherz-Eichinger, D.; Erdoes, G.; Berlakovich, G.; Bacher, A.; Krenn, C. G.; Faybik, P. Alterations of Endothelial Glycocalyx During Orthotopic Liver Transplantation in Patients With End-Stage Liver Disease. *Transplantation* **2015**, *99*, 2118–2123, doi:10.1097/TP.0000000000000680.

34. Richter, V.; Savery, M. D.; Gassmann, M.; Baum, O.; Damiano, E. R.; Pries, A. R. Excessive erythrocytosis compromises the blood-endothelium interface in erythropoietin-overexpressing mice. *J. Physiol.* **2011**, *589*, 5181–5192, doi:10.1113/jphysiol.2011.209262.

35. Nijst, P.; Cops, J.; Martens, P.; Swennen, Q.; Dupont, M.; Tang, W. H. W.; Mullens, W. Endovascular shedding markers in patients with heart failure with reduced ejection fraction: Results from a single-center exploratory study. *Microcirc. N. Y. N 1994* **2018**, *25*, doi:10.1111/micc.12432.

36. Martens, R. J. H.; Vink, H.; van Oostenbrugge, R. J.; Staals, J. Sublingual microvascular glycocalyx dimensions in lacunar stroke patients. *Cerebrovasc. Dis. Basel Switz.* **2013**, *35*, 451–454, doi:10.1159/000348854.

37. Yamazaki, Y.; Baker, D. J.; Tachibana, M.; Liu, C.-C.; van Deursen, J. M.; Brott, T. G.; Bu, G.; Kanekiyo, T. Vascular Cell Senescence Contributes to Blood-Brain Barrier Breakdown. *Stroke* **2016**, *47*, 1068–1077, doi:10.1161/STROKEAHA.115.010835.

38. Dias, H. K. I.; Brown, C. L. R.; Polidori, M. C.; Lip, G. Y. H.; Griffiths, H. R. LDL-lipids from patients with hypercholesterolaemia and Alzheimer's disease are inflammatory to microvascular endothelial cells: mitigation by statin intervention. *Clin. Sci. Lond. Engl. 1979* **2015**, *129*, 1195–1206, doi:10.1042/CS20150351.

39. Wang, L.; Chen, Y.; Li, X.; Zhang, Y.; Gulbins, E.; Zhang, Y. Enhancement of endothelial permeability by free fatty acid through lysosomal cathepsin B-mediated Nlrp3 inflammasome activation. *Oncotarget* **2016**, *7*, 73229–73241, doi:10.18632/oncotarget.12302.

40. Haidari, M.; Zhang, W.; Willerson, J. T.; Dixon, R. A. Disruption of endothelial adherens junctions by high glucose is mediated by protein kinase C-β-dependent vascular endothelial cadherin tyrosine phosphorylation. *Cardiovasc. Diabetol.* **2014**, *13*, 105, doi:10.1186/1475-2840-13-105.

41. Hernandez, N. M.; Casselbrant, A.; Joshi, M.; Johansson, B. R.; Sumitran-Holgersson, S. Antibodies to kidney endothelial cells contribute to a "leaky" glomerular barrier in patients with chronic kidney diseases. *Am. J. Physiol. Renal Physiol.* **2012**, *302*, F884-894, doi:10.1152/ajprenal.00250.2011.

42. Lu, Q.; Gottlieb, E.; Rounds, S. Effects of Cigarette Smoke on Pulmonary Endothelial Cells. *Am. J. Physiol. Lung Cell. Mol. Physiol.* **2018**, doi:10.1152/ajplung.00373.2017.

43. Polverino, F.; Celli, B. R.; Owen, C. A. COPD as an endothelial disorder: endothelial injury linking lesions in the lungs and other organs? (2017 Grover Conference Series). *Pulm. Circ.* **2018**, *8*, 2045894018758528, doi:10.1177/2045894018758528.

44. Cerutti, C.; Ridley, A. J. Endothelial cell-cell adhesion and signaling. *Exp. Cell Res.* **2017**, *358*, 31–38, doi:10.1016/j.yexcr.2017.06.003.

45. Romanov, Y. A.; Chervontseva, A. M.; Savchenko, V. G.; Smirnov, V. N. Vascular endothelium: target or victim of cytostatic therapy? *Can. J. Physiol. Pharmacol.* **2007**, *85*, 396–403, doi:10.1139/y07-045.

46. Guipaud, O.; Jaillet, C.; Clément-Colmou, K.; François, A.; Supiot, S.; Milliat, F. The importance of the vascular endothelial barrier in the immune-inflammatory response induced by radiotherapy. *Br. J. Radiol.* **2018**, 20170762, doi:10.1259/bjr.20170762.

47. Chistiakov, D. A.; Orekhov, A. N.; Bobryshev, Y. V. Endothelial Barrier and Its Abnormalities in Cardiovascular Disease. *Front. Physiol.* **2015**, *6*, 365, doi:10.3389/fphys.2015.00365.

48. Ferrer, R.; Mateu, X.; Maseda, E.; Yébenes, J. C.; Aldecoa, C.; De Haro, C.; Ruiz-Rodriguez, J. C.; Garnacho-Montero, J. Non-oncotic properties of albumin. A multidisciplinary vision about the implications for critically ill patients. *Expert Rev. Clin. Pharmacol.* **2018**, *11*, 125–137, doi:10.1080/17512433.2018.1412827.

49. Tarbell, J. M.; Cancel, L. M. The glycocalyx and its significance in human medicine. *J. Intern. Med.* **2016**, *280*, 97–113, doi:10.1111/joim.12465.

50. Huynh, J.; Nishimura, N.; Rana, K.; Peloquin, J. M.; Califano, J. P.; Montague, C. R.; King, M. R.; Schaffer, C. B.; Reinhart-King, C. A. Age-related intimal stiffening enhances endothelial permeability and leukocyte transmigration. *Sci. Transl. Med.* **2011**, *3*, 112ra122, doi:10.1126/scitranslmed.3002761.

51. Sena, C. M.; Pereira, A. M.; Seiça, R. Endothelial dysfunction - a major mediator of diabetic vascular disease. *Biochim. Biophys. Acta* **2013**, *1832*, 2216–2231, doi:10.1016/j.bbadis.2013.08.006.

52. Fu, Q.; Colgan, S. P.; Shelley, C. S. Hypoxia: The Force that Drives Chronic Kidney Disease. *Clin. Med. Res.* **2016**, *14*, 15–39, doi:10.3121/cmr.2015.1282.

53. Favero, G.; Paganelli, C.; Buffoli, B.; Rodella, L. F.; Rezzani, R. Endothelium and its alterations in cardiovascular diseases: life style intervention. *BioMed Res. Int.* **2014**, *2014*, 801896, doi:10.1155/2014/801896.

54. Wadkin, J. C. R.; Patten, D. A.; Kamarajah, S. K.; Shepherd, E. L.; Novitskaya, V.; Berditchevski, F.; Adams, D. H.; Weston, C. J.; Shetty, S. CD151 supports VCAM-1-mediated lymphocyte adhesion to liver endothelium and is upregulated in chronic liver disease and hepatocellular carcinoma. *Am. J. Physiol. Gastrointest. Liver Physiol.* **2017**, *313*, G138–G149, doi:10.1152/ajpgi.00411.2016.

55. Gavin, J. B.; Maxwell, L.; Edgar, S. G. Microvascular involvement in cardiac pathology. *J. Mol. Cell. Cardiol.* **1998**, *30*, 2531–2540, doi:10.1006/jmcc.1998.0824.

56. Frijns, C. J. M.; Kappelle, L. J. Inflammatory cell adhesion molecules in ischemic cerebrovascular disease. *Stroke* **2002**, *33*, 2115–2122.

57. Silva, G. C.; Abbas, M.; Khemais-Benkhiat, S.; Burban, M.; Ribeiro, T. P.; Toti, F.; Idris-Khodja, N.; Côrtes, S. F.; Schini-Kerth, V. B. Replicative senescence promotes prothrombotic responses in endothelial cells: Role of NADPH oxidase- and cyclooxygenase-derived oxidative stress. *Exp. Gerontol.* **2017**, *93*, 7–15, doi:10.1016/j.exger.2017.04.006.

58. Brandes, R. P. Endothelial dysfunction and hypertension. *Hypertens. Dallas Tex 1979* **2014**, *64*, 924–928, doi:10.1161/HYPERTENSIONAHA.114.03575.

59. Lutz, J.; Menke, J.; Sollinger, D.; Schinzel, H.; Thürmel, K. Haemostasis in chronic kidney disease. *Nephrol. Dial. Transplant. Off. Publ. Eur. Dial. Transpl. Assoc. - Eur. Ren. Assoc.* **2014**, *29*, 29–40, doi:10.1093/ndt/gft209.

60. Levi, M. Cancer-related coagulopathies. *Thromb. Res.* **2014**, *133 Suppl 2*, S70-75, doi:10.1016/S0049-3848(14)50012-6.

61. Remková, A.; Remko, M. Homocysteine and endothelial markers are increased in patients with chronic liver diseases. *Eur. J. Intern. Med.* **2009**, *20*, 482–486, doi:10.1016/j.ejim.2009.03.002.

62. Besedina, A. NO-Synthase Activity in Patients with Coronary Heart Disease Associated with Hypertension of Different Age Groups. *J. Med. Biochem.* **2016**, *35*, 43–49, doi:10.1515/jomb-2015-0008.

63. Wiseman, S.; Marlborough, F.; Doubal, F.; Webb, D. J.; Wardlaw, J. Blood markers of coagulation, fibrinolysis, endothelial dysfunction and inflammation in lacunar stroke versus non-lacunar stroke and non-stroke: systematic review and meta-analysis. *Cerebrovasc. Dis. Basel Switz.* **2014**, *37*, 64–75, doi:10.1159/000356789.

64. Kung, C.-T.; Su, C.-M.; Chen, C. T.; Cheng, H.-H.; Chang, M.-W.; Hung, C.-W.; Hung, S.-C.; Chang, W.-N.; Tsai, N.-W.; Wang, H.-C.; Su, Y.-J.; Huang, C.-C.; Lin, W.-C.; Cheng, B.-C.; Chang, Y.-T.; Lu, C.-H. Circulating endothelial progenitor cells may predict outcomes in adult patients with severe sepsis in the emergency department. *Clin. Chim. Acta Int. J. Clin. Chem.* **2016**, *455*, 1–6, doi:10.1016/j.cca.2016.01.015.

65. Williamson, K. A.; Hamilton, A.; Reynolds, J. A.; Sipos, P.; Crocker, I.; Stringer, S. E.; Alexander, Y. M. Age-related impairment of endothelial progenitor cell migration correlates with structural alterations of heparan sulfate proteoglycans. *Aging Cell* **2013**, *12*, 139–147, doi:10.1111/acel.12031.

66. Burger, D.; Touyz, R. M. Cellular biomarkers of endothelial health: microparticles, endothelial progenitor cells, and circulating endothelial cells. *J. Am. Soc. Hypertens. JASH* **2012**, *6*, 85–99, doi:10.1016/j.jash.2011.11.003.

67. Paschalaki, K. E.; Starke, R. D.; Hu, Y.; Mercado, N.; Margariti, A.; Gorgoulis, V. G.; Randi, A. M.; Barnes, P. J. Dysfunction of endothelial progenitor cells from smokers and chronic obstructive pulmonary disease patients due to increased DNA damage and senescence. *Stem Cells Dayt. Ohio* **2013**, *31*, 2813–2826, doi:10.1002/stem.1488.

68. Pradhan, K.; Mund, J.; Case, J.; Gupta, S.; Liu, Z.; Gathirua-Mwangi, W.; McDaniel, A.; Renbarger, J.; Champion, V. Differences in Circulating Endothelial Progenitor Cells among Childhood Cancer Survivors Treated with and without Radiation. *J. Hematol. Thromb.* **2015**, *1*, doi:10.13188/2380-6842.1000005.

69. Wang, J.; Wu, A.; Wu, Y. Endothelial Glycocalyx Layer: A Possible Therapeutic Target for Acute Lung Injury during Lung Resection. *BioMed Res. Int.* **2017**, *2017*, 5969657, doi:10.1155/2017/5969657.

70. Chappell, D.; Bruegger, D.; Potzel, J.; Jacob, M.; Brettner, F.; Vogeser, M.; Conzen, P.; Becker, B. F.; Rehm, M. Hypervolemia increases release of atrial natriuretic peptide and shedding of the endothelial glycocalyx. *Crit. Care Lond. Engl.* **2014**, *18*, 538, doi:10.1186/s13054-014-0538-5.